\documentclass{article}

\usepackage[final]{neurips_2023}

\usepackage[utf8]{inputenc} 
\usepackage[T1]{fontenc}    
\usepackage{hyperref}       
\usepackage{url}            
\usepackage{booktabs}       
\usepackage{amsfonts}       
\usepackage{nicefrac}       
\usepackage{microtype}      
\usepackage{xcolor}         

\title{Beyond Fairness: Alternative Moral Dimensions for Assessing Algorithms and Designing Systems}

\author{%
  Kimi Wenzel
  \\
  Human-Computer Interaction Institute\\
  Carnegie Mellon University\\
  Pittsburgh, PA 15213 \\
  \texttt{kwenzel@cs.cmu.edu} \\
   \And
   Geoff Kaufman \\
   Human-Computer Interaction Institute\\
   Carnegie Mellon University \\
   Pittsburgh, PA 15213 \\
   \texttt{gfk@cs.cmu.edu} \\
   \AND
   Laura Dabbish \\
   Human-Computer Interaction Institute\\
   Carnegie Mellon University \\
   Pittsburgh, PA 15213 \\
   \texttt{dabbish@cmu.edu} \\
}

\begin{document}

\maketitle

\begin{abstract}
The ethics of artificial intelligence (AI) systems has risen as an imminent concern across scholarly communities. This concern has propagated a great interest in algorithmic \textit{fairness}. Large research agendas are now devoted to increasing algorithmic fairness, assessing algorithmic fairness, and understanding human perceptions of fairness. We argue that there is an overreliance on fairness as a single dimension of morality, which comes at the expense of other important human values. Drawing from moral psychology, we present five moral dimensions that go beyond fairness, and suggest three ways these alternative dimensions may contribute to ethical AI development. 
\end{abstract}

\section{Introduction}


Evaluations of morality in algorithmic systems typically take a \textit{monist} approach, optimizing for a single dimension of morality, namely, fairness \citep{mitchell2021algorithmic}. In contrast, models of moral reasoning have proposed a multitude of moral pillars, in addition to fairness, that guide human judgment and behavior. The following paper describes factors contributing to the positioning of fairness as a central focus in the evaluation of algorithms, provides a review of alternative moral values derived from psychological theories, and finally makes a case for how expanding our descriptive understanding of morality can enhance the ethics of technologists and help design algorithmic tools with a broad user acceptance. 

\section{The Reign of Algorithmic Fairness}
Why has fairness prevailed over other moral dimensions? There are three factors we describe here: (1) the prioritization of Western ideals, (2) the context of deployment, and (3) the mathematical ``messiness''\footnote{See Graham et al.'s (p. 57) summary of William James's work from 1909 \citep{graham2013moral}.} of morality. We also describe unique problems that arise for each factor, respectively.

\subsection{Prioritization of Western Ideals}
Fairness is a highly weighted virtue in the West \citep{sambasivan2021re, velasquez1990justice, dator2006fairness}. Many non-Western cultures also value fairness; however, in such contexts fairness often coexists with many other virtues \citep{shweder2013big, dator2006fairness, graham2011mapping}, which in turn affects how ``fair'' algorithmic systems are deployed and accepted \citep{sambasivan2021re}. Thus, when evaluating the descriptive morality of algorithmic systems, it is important then to not focus solely on fairness, as has been the precedent for several years now \citep{starke2022fairness}, but to evaluate multiple dimensions of morality. 

\subsection{Context of Deployment}
While fairness as it pertains to mathematics and algorithms has been studied for centuries, its recent popularity has largely been towards the development of risk assessments for legal systems \citep{longhistory2020, eubanks2018automating}. This legal context, even in the absence of algorithmic tools, aims to provide fair outcomes. Thus, when building tools for these types of risk assessments, optimizing for algorithmic fairness can appear logical. This simplistic framing however, can promote broader contextual unawareness and ultimately fails to acknowledge the social, environmental, and economic structures that contribute to unfair and unethical outcomes \citep{hoffmann2019fairness, selbst2019fairness}. Beyond basic contextual awareness, another problem has arisen regarding how the context of fairness applications itself has changed. Fairness's scholarly boom has led it to be adapted and applied to various domains, sometimes with insufficient reasoning justifying its application. For example, fairness may have inappropriately superceded \textit{liberty} in the context of biometrics \citep{valdivia2022there} and superceded \textit{harm} in automatic speech recognition \citep{wenzel2023can}. The latter example demonstrated how fair and equal error rates in voice assistant interactions can lead to disparate mental health outcomes across races, demonstrating how championing fairness can fail to the extent that it ignores the pre-existing social landscape (i.e. racism), and it can also fail to the extent that it undermines other values (i.e. \textit{care}). (See Table \ref{MFTtable} for definitions.) 



\subsection{Mathematical ``Messiness'' of Morality} \label{sec:math}

Fairness, in spite of the difficulties associated with defining and operationalizing it \citep{mitchell2021algorithmic}, may be perceived as more compatible with mathematics. In his article about fairness and political philosophy, Binns describes how ```fairness' as used in the fair machine learning community is best understood as a placeholder term for a variety
of normative egalitarian considerations'' \citep{binns2018fairness}. This egalitarian interpretation held by computer scientists points to how fairness may be, in its most essential form, composed of equality conditions. For example, a model for facial recognition may be considered fair if the error rates are equal across all races. Other moral dimensions have fewer a priori mappings to equations and thus are more difficult to operationalize, perhaps due to the social intuition involved in moral judgements \citep{haidt2001emotional}. Adding to the complexity, embracing moral pluralism would require not only translating moral values into optimization problems, but would also require that all these moral values are respected and upheld. Considering how pluralistic applications of mathematical fairness alone are impossible \citep{kleinberg2016inherent}, there is probable concern that pluralistic applications of morality more broadly may pose similar problems.

\section{Taking a Step Forward}
Built upon years of prior work, the Moral Foundations Theory (MFT) differentiates various facets of morality, namely \textit{care}, \textit{fairness}, \textit{loyalty}, \textit{authority}, and \textit{sanctity} \citep{graham2013moral}, with some research arguing to include other dimensions such as \textit{liberty} \citep{haidt2012righteous}; see Table \ref{MFTtable}. MFT is perhaps most known for explaining differences in moral judgement between conservatives and liberals in the United States \citep{haidt2012righteous, haidt2007morality, graham2009liberals}; however, it has been adapted for non-WEIRD countries (i.e. non- Western, educated, industrialized, rich, democratic countries) \citep{dougruyol2019five, atari2022morality, matsuo2019development} and has been studied in contexts as diverse as the life aspirations of Mongolian youth \citep{bespalov2017life}, nuclear disarmament in South Africa \citep{das2022role}, and Islamic business in Russia \citep{penikas2021textual}. We propose using MFT to reframe research on AI ethics. Understanding human perceptions of morality, beyond fairness, may (a) help us understand user adoption of algorithmic systems, (b) help us understand how to best build interventions for technologists creating systems with algorithmic power, and (c) orient systems away from technosolutionism and toward society-in-the-loop methods. 

\begin{table}[h]
  \caption{Six moral foundations, adapted from \citep{grizzard2017morality, haidt2012righteous}}
  \label{MFTtable}
  \centering
  \begin{tabular}{lll}
    \toprule
    Foundation     & Upholding actions    \\
    \midrule
    Care & Alleviate physical or mental suffering     \\
    Fairness     & Foster equality      \\
    Loyalty     & Act loyal to one’s ingroup   \\
    Authority & Respect relationships within hierarchies     \\
    Sanctity     & Avoid physical, mental, or spiritual contamination      \\
    Liberty     & Protect freedom; Keep dominant agents ``in-check''       \\
    \bottomrule
  \end{tabular}
\end{table}

\paragraph{a. Promote User Adoption} Aligning a product's ethical values with users' ethical values using MFT may help boost user acceptance \citep{telkamp2022implications}. Thus, MFT may not only help improve the ethical integrity of algorithmic systems, but may also help improve user satisfaction more broadly. Building upon this, MFT may also help explain differences in perceptions of AI across different cultures. For example, might higher valuations of \textit{authority}, \textit{loyality}, or \textit{sanctity} among Indians, vs. Americans and Europeans, \citep{renner2014globalization} explain the observed aspiration toward AI among Indian nationals \citep{sambasivan2021re}? Such insights may help design more culturally sensitive algorithmic systems. We encourage further research be done to evaluate how existing algorithmic systems may be perceived by users, based on their MFT orientations.



\paragraph{b. Build Interventions} MFT has been shown to help facilitate collective action \citep{milesi2018pluralistic}. Thus, it may be worth investigating how we may harness MFT to help facilitate collective action toward ethical goals in computing. Respect for individuals' MFT orientations can also facilitate better cross-cultural collaboration \citep{romig2018using}, so there is sufficient reason to believe studying diverse users' MFT orientations could assist researchers' efforts in creating technologies for cooperative work.

\paragraph{c. Facilitate a Social Contract} The mathematical difficulties (Sec \ref{sec:math}) associated with deploying diverse MFT foundations may actually be beneficial to the extent that they discourage technosolutionism: If a set of moral foundations cannot be readily optimized in an algorithmic system, designers and developers may consider alternative solutions that integrate human feedback and oversight into the product lifecycle. Such solutions may harness human-in-the-loop (HITL) and human-AI complementarity methods, which have been recommended by researchers for AI tools used in education \citep{holstein2019designing}, social services \citep{kawakami2022improving, fogliato2022case}, and healthcare \citep{sivaraman2023ignore}. These methods afford users the ability to integrate their own expertise and moral oversight into algorithmic decisions. Going a step past HITL methods, we encourage a society-in-the-loop (SITL) approach. SITL builds upon HITL by integrating a \textit{social contract} into the cycle \citep{rahwan2018society}. Importantly, the social contract includes groups with conflicting interests and values. Translating these values to MFT foundations, and integrating MFT foundations into the SITL method may support a more comprehensive social contract and a more inclusive values system. 





\section{Limitations} As we work to embed a diverse set of morals into emerging technologies, we must think critically about whose values will be prioritized. Importantly, MFT is a descriptive theory that makes sense of people's moral judgements -- it is not a normative theory intended to guide human behavior. Should we be designing technologies to match users' moral orientations, so as to best promote user acceptance? Should we leverage persuasive design techniques to re-orient individuals towards a normative ethic? We must approach this research agenda with great care and intentionality, knowing that appealing too far towards users' moral biases can cause harm \citep{ranalli2023s}, and trying to persuade users can sometimes accelerate abuse \citep{hunter2016facebook}. 

To this point, we stress that any integration of MFT should be \textit{culturally situated}. While MFT alone does not have norm embeddings, it may still inform Norm Sensitive Design, and help adapt value-sensitive methodologies to diverse cultural contexts \citep{martin2023we}. Furthermore, we also advocate for \textit{full integration} of diverse and historically marginalized technologists into the AI product lifecycle. Importantly, this integration goes beyond simply having marginalized identities represented in teams, but also ensuring that they are actively engaged and uplifted \citep{smith2017diversity}. Adequate integration helps ensure that algorithmic systems respect pluralistic values.

\section{Conclusion}

As algorithms are increasingly deployed in both high-stakes and everyday consumer contexts, it's imperative that they are not exclusively optimized for fairness, but rather, are designed to uphold the ethical integrity of our diverse and morally pluralistic society. In this paper, we have outlined three reasons driving fairness's popularity over other moral values (prioritization of Western ideals, the context of legal risk assessments, mathematical challenges) and describe challenges associated with each reason. We then introduced Haidt's Moral Foundation's Theory and described how it may be adapted to computing research by promoting user adoption, informing ethics interventions, and assisting in the development of important social contracts. Lastly, we stress the value of diversity and cultural awareness in the development of AI systems. Going forward, we encourage research studying human perceptions of fairness in algorithmic systems expands to include human perceptions of \textit{morality} more broadly. If a core tenent of responsible AI is to ``align system goals with human values'' \citep{dignum2019responsible}, we urge technologists to continually question whose human values they are aligning to, and whose values are being ignored.


\bibliographystyle{ACM-Reference-Format}
\bibliography{references.bib}


\begin{thebibliography}{41}


\ifx \showCODEN    \undefined \def \showCODEN     #1{\unskip}     \fi
\ifx \showDOI      \undefined \def \showDOI       #1{#1}\fi
\ifx \showISBNx    \undefined \def \showISBNx     #1{\unskip}     \fi
\ifx \showISBNxiii \undefined \def \showISBNxiii  #1{\unskip}     \fi
\ifx \showISSN     \undefined \def \showISSN      #1{\unskip}     \fi
\ifx \showLCCN     \undefined \def \showLCCN      #1{\unskip}     \fi
\ifx \shownote     \undefined \def \shownote      #1{#1}          \fi
\ifx \showarticletitle \undefined \def \showarticletitle #1{#1}   \fi
\ifx \showURL      \undefined \def \showURL       {\relax}        \fi
\providecommand\bibfield[2]{#2}
\providecommand\bibinfo[2]{#2}
\providecommand\natexlab[1]{#1}
\providecommand\showeprint[2][]{arXiv:#2}

\bibitem[Atari et~al\mbox{.}(2022)]%
        {atari2022morality}
\bibfield{author}{\bibinfo{person}{Mohammad Atari}, \bibinfo{person}{Jonathan Haidt}, \bibinfo{person}{Jesse Graham}, \bibinfo{person}{Sena Koleva}, \bibinfo{person}{Sean~T Stevens}, {and} \bibinfo{person}{Morteza Dehghani}.} \bibinfo{year}{2022}\natexlab{}.
\newblock \showarticletitle{Morality beyond the WEIRD: How the nomological network of morality varies across cultures}.
\newblock  (\bibinfo{year}{2022}).
\newblock


\bibitem[Bespalov et~al\mbox{.}(2017)]%
        {bespalov2017life}
\bibfield{author}{\bibinfo{person}{Alexander Bespalov}, \bibinfo{person}{Marina Prudnikova}, \bibinfo{person}{Bavuu Nyamdorj}, {and} \bibinfo{person}{Mikhail Vlasov}.} \bibinfo{year}{2017}\natexlab{}.
\newblock \showarticletitle{Life aspirations, values and moral foundations in Mongolian youth}.
\newblock \bibinfo{journal}{\emph{Journal of Moral Education}} \bibinfo{volume}{46}, \bibinfo{number}{3} (\bibinfo{year}{2017}), \bibinfo{pages}{258--271}.
\newblock


\bibitem[Binns(2018)]%
        {binns2018fairness}
\bibfield{author}{\bibinfo{person}{Reuben Binns}.} \bibinfo{year}{2018}\natexlab{}.
\newblock \showarticletitle{Fairness in machine learning: Lessons from political philosophy}. In \bibinfo{booktitle}{\emph{Conference on fairness, accountability and transparency}}. PMLR, \bibinfo{pages}{149--159}.
\newblock


\bibitem[Das(2022)]%
        {das2022role}
\bibfield{author}{\bibinfo{person}{Arunjana Das}.} \bibinfo{year}{2022}\natexlab{}.
\newblock \showarticletitle{Role of moral foundations in the nuclear disarmament of South Africa}.
\newblock \bibinfo{journal}{\emph{Scientia Militaria: South African Journal of Military Studies}} \bibinfo{volume}{50}, \bibinfo{number}{1} (\bibinfo{year}{2022}), \bibinfo{pages}{91--119}.
\newblock


\bibitem[Dator et~al\mbox{.}(2006)]%
        {dator2006fairness}
\bibfield{author}{\bibinfo{person}{Jim Dator}, \bibinfo{person}{D Pratt}, {and} \bibinfo{person}{Y Seo}.} \bibinfo{year}{2006}\natexlab{}.
\newblock \showarticletitle{What Is Fairness}.
\newblock \bibinfo{journal}{\emph{Fairness, Globalization, and Public Institutions}} (\bibinfo{year}{2006}), \bibinfo{pages}{19}.
\newblock


\bibitem[Dignum(2019)]%
        {dignum2019responsible}
\bibfield{author}{\bibinfo{person}{Virginia Dignum}.} \bibinfo{year}{2019}\natexlab{}.
\newblock \bibinfo{booktitle}{\emph{Responsible artificial intelligence: how to develop and use AI in a responsible way}}. Vol.~\bibinfo{volume}{2156}.
\newblock \bibinfo{publisher}{Springer}.
\newblock


\bibitem[Do{\u{g}}ruyol et~al\mbox{.}(2019)]%
        {dougruyol2019five}
\bibfield{author}{\bibinfo{person}{Burak Do{\u{g}}ruyol}, \bibinfo{person}{Sinan Alper}, {and} \bibinfo{person}{Onurcan Yilmaz}.} \bibinfo{year}{2019}\natexlab{}.
\newblock \showarticletitle{The five-factor model of the moral foundations theory is stable across WEIRD and non-WEIRD cultures}.
\newblock \bibinfo{journal}{\emph{Personality and Individual Differences}}  \bibinfo{volume}{151} (\bibinfo{year}{2019}), \bibinfo{pages}{109547}.
\newblock


\bibitem[Eubanks(2018)]%
        {eubanks2018automating}
\bibfield{author}{\bibinfo{person}{Virginia Eubanks}.} \bibinfo{year}{2018}\natexlab{}.
\newblock \bibinfo{booktitle}{\emph{Automating inequality: How high-tech tools profile, police, and punish the poor}}.
\newblock \bibinfo{publisher}{St. Martin's Press}.
\newblock


\bibitem[Fogliato et~al\mbox{.}(2022)]%
        {fogliato2022case}
\bibfield{author}{\bibinfo{person}{Riccardo Fogliato}, \bibinfo{person}{Maria De-Arteaga}, {and} \bibinfo{person}{Alexandra Chouldechova}.} \bibinfo{year}{2022}\natexlab{}.
\newblock \showarticletitle{A case for humans-in-the-loop: Decisions in the presence of misestimated algorithmic scores}.
\newblock \bibinfo{journal}{\emph{Available at SSRN}} (\bibinfo{year}{2022}).
\newblock


\bibitem[Graham et~al\mbox{.}(2013)]%
        {graham2013moral}
\bibfield{author}{\bibinfo{person}{Jesse Graham}, \bibinfo{person}{Jonathan Haidt}, \bibinfo{person}{Sena Koleva}, \bibinfo{person}{Matt Motyl}, \bibinfo{person}{Ravi Iyer}, \bibinfo{person}{Sean~P Wojcik}, {and} \bibinfo{person}{Peter~H Ditto}.} \bibinfo{year}{2013}\natexlab{}.
\newblock \showarticletitle{Moral foundations theory: The pragmatic validity of moral pluralism}.
\newblock In \bibinfo{booktitle}{\emph{Advances in experimental social psychology}}. Vol.~\bibinfo{volume}{47}. \bibinfo{publisher}{Elsevier}, \bibinfo{pages}{55--130}.
\newblock


\bibitem[Graham et~al\mbox{.}(2009)]%
        {graham2009liberals}
\bibfield{author}{\bibinfo{person}{Jesse Graham}, \bibinfo{person}{Jonathan Haidt}, {and} \bibinfo{person}{Brian~A Nosek}.} \bibinfo{year}{2009}\natexlab{}.
\newblock \showarticletitle{Liberals and conservatives rely on different sets of moral foundations.}
\newblock \bibinfo{journal}{\emph{Journal of personality and social psychology}} \bibinfo{volume}{96}, \bibinfo{number}{5} (\bibinfo{year}{2009}), \bibinfo{pages}{1029}.
\newblock


\bibitem[Graham et~al\mbox{.}(2011)]%
        {graham2011mapping}
\bibfield{author}{\bibinfo{person}{Jesse Graham}, \bibinfo{person}{Brian~A Nosek}, \bibinfo{person}{Jonathan Haidt}, \bibinfo{person}{Ravi Iyer}, \bibinfo{person}{Spassena Koleva}, {and} \bibinfo{person}{Peter~H Ditto}.} \bibinfo{year}{2011}\natexlab{}.
\newblock \showarticletitle{Mapping the moral domain.}
\newblock \bibinfo{journal}{\emph{Journal of personality and social psychology}} \bibinfo{volume}{101}, \bibinfo{number}{2} (\bibinfo{year}{2011}), \bibinfo{pages}{366}.
\newblock


\bibitem[Grizzard and Ahn(2017)]%
        {grizzard2017morality}
\bibfield{author}{\bibinfo{person}{Matthew Grizzard} {and} \bibinfo{person}{Changhyun Ahn}.} \bibinfo{year}{2017}\natexlab{}.
\newblock \showarticletitle{Morality \& personality: perfect and deviant selves}.
\newblock \bibinfo{journal}{\emph{Avatar, Assembled: The Social and Technical Anatomy of Digital Bodies. New York, NY: Peter Lang}} (\bibinfo{year}{2017}).
\newblock


\bibitem[Haidt(2001)]%
        {haidt2001emotional}
\bibfield{author}{\bibinfo{person}{Jonathan Haidt}.} \bibinfo{year}{2001}\natexlab{}.
\newblock \showarticletitle{The emotional dog and its rational tail: a social intuitionist approach to moral judgment.}
\newblock \bibinfo{journal}{\emph{Psychological review}} \bibinfo{volume}{108}, \bibinfo{number}{4} (\bibinfo{year}{2001}), \bibinfo{pages}{814}.
\newblock


\bibitem[Haidt(2012)]%
        {haidt2012righteous}
\bibfield{author}{\bibinfo{person}{Jonathan Haidt}.} \bibinfo{year}{2012}\natexlab{}.
\newblock \bibinfo{booktitle}{\emph{The righteous mind: Why good people are divided by politics and religion}}.
\newblock \bibinfo{publisher}{Vintage}.
\newblock


\bibitem[Haidt and Graham(2007)]%
        {haidt2007morality}
\bibfield{author}{\bibinfo{person}{Jonathan Haidt} {and} \bibinfo{person}{Jesse Graham}.} \bibinfo{year}{2007}\natexlab{}.
\newblock \showarticletitle{When morality opposes justice: Conservatives have moral intuitions that liberals may not recognize}.
\newblock \bibinfo{journal}{\emph{Social justice research}} \bibinfo{volume}{20}, \bibinfo{number}{1} (\bibinfo{year}{2007}), \bibinfo{pages}{98--116}.
\newblock


\bibitem[Hoffmann(2019)]%
        {hoffmann2019fairness}
\bibfield{author}{\bibinfo{person}{Anna~Lauren Hoffmann}.} \bibinfo{year}{2019}\natexlab{}.
\newblock \showarticletitle{Where fairness fails: data, algorithms, and the limits of antidiscrimination discourse}.
\newblock \bibinfo{journal}{\emph{Information, Communication \& Society}} \bibinfo{volume}{22}, \bibinfo{number}{7} (\bibinfo{year}{2019}), \bibinfo{pages}{900--915}.
\newblock


\bibitem[Holstein et~al\mbox{.}(2019)]%
        {holstein2019designing}
\bibfield{author}{\bibinfo{person}{Kenneth Holstein}, \bibinfo{person}{Bruce~M McLaren}, {and} \bibinfo{person}{Vincent Aleven}.} \bibinfo{year}{2019}\natexlab{}.
\newblock \showarticletitle{Designing for complementarity: Teacher and student needs for orchestration support in AI-enhanced classrooms}. In \bibinfo{booktitle}{\emph{Artificial Intelligence in Education: 20th International Conference, AIED 2019, Chicago, IL, USA, June 25-29, 2019, Proceedings, Part I 20}}. Springer, \bibinfo{pages}{157--171}.
\newblock


\bibitem[Hunter and Evans(2016)]%
        {hunter2016facebook}
\bibfield{author}{\bibinfo{person}{David Hunter} {and} \bibinfo{person}{Nicholas Evans}.} \bibinfo{year}{2016}\natexlab{}.
\newblock \bibinfo{title}{Facebook emotional contagion experiment controversy}.
\newblock , \bibinfo{numpages}{2--3}~pages.
\newblock


\bibitem[Kawakami et~al\mbox{.}(2022)]%
        {kawakami2022improving}
\bibfield{author}{\bibinfo{person}{Anna Kawakami}, \bibinfo{person}{Venkatesh Sivaraman}, \bibinfo{person}{Hao-Fei Cheng}, \bibinfo{person}{Logan Stapleton}, \bibinfo{person}{Yanghuidi Cheng}, \bibinfo{person}{Diana Qing}, \bibinfo{person}{Adam Perer}, \bibinfo{person}{Zhiwei~Steven Wu}, \bibinfo{person}{Haiyi Zhu}, {and} \bibinfo{person}{Kenneth Holstein}.} \bibinfo{year}{2022}\natexlab{}.
\newblock \showarticletitle{Improving human-AI partnerships in child welfare: understanding worker practices, challenges, and desires for algorithmic decision support}. In \bibinfo{booktitle}{\emph{Proceedings of the 2022 CHI Conference on Human Factors in Computing Systems}}. \bibinfo{pages}{1--18}.
\newblock


\bibitem[Kleinberg et~al\mbox{.}(2016)]%
        {kleinberg2016inherent}
\bibfield{author}{\bibinfo{person}{Jon Kleinberg}, \bibinfo{person}{Sendhil Mullainathan}, {and} \bibinfo{person}{Manish Raghavan}.} \bibinfo{year}{2016}\natexlab{}.
\newblock \showarticletitle{Inherent trade-offs in the fair determination of risk scores}.
\newblock \bibinfo{journal}{\emph{arXiv preprint arXiv:1609.05807}} (\bibinfo{year}{2016}).
\newblock


\bibitem[Martin et~al\mbox{.}(2023)]%
        {martin2023we}
\bibfield{author}{\bibinfo{person}{Diana~Adela Martin}, \bibinfo{person}{Rockwell~F Clancy}, \bibinfo{person}{Qin Zhu}, {and} \bibinfo{person}{Gunter Bombaerts}.} \bibinfo{year}{2023}\natexlab{}.
\newblock \showarticletitle{Why do we need Norm Sensitive Design? A WEIRD critique of value sensitive approaches to design}.
\newblock  (\bibinfo{year}{2023}).
\newblock


\bibitem[Matsuo et~al\mbox{.}(2019)]%
        {matsuo2019development}
\bibfield{author}{\bibinfo{person}{Akiko Matsuo}, \bibinfo{person}{Kazutoshi Sasahara}, \bibinfo{person}{Yasuhiro Taguchi}, {and} \bibinfo{person}{Minoru Karasawa}.} \bibinfo{year}{2019}\natexlab{}.
\newblock \showarticletitle{Development and validation of the Japanese moral foundations dictionary}.
\newblock \bibinfo{journal}{\emph{PloS one}} \bibinfo{volume}{14}, \bibinfo{number}{3} (\bibinfo{year}{2019}), \bibinfo{pages}{e0213343}.
\newblock


\bibitem[Milesi and Alberici(2018)]%
        {milesi2018pluralistic}
\bibfield{author}{\bibinfo{person}{Patrizia Milesi} {and} \bibinfo{person}{Augusta~Isabella Alberici}.} \bibinfo{year}{2018}\natexlab{}.
\newblock \showarticletitle{Pluralistic morality and collective action: The role of moral foundations}.
\newblock \bibinfo{journal}{\emph{Group Processes \& Intergroup Relations}} \bibinfo{volume}{21}, \bibinfo{number}{2} (\bibinfo{year}{2018}), \bibinfo{pages}{235--256}.
\newblock


\bibitem[Mitchell et~al\mbox{.}(2021)]%
        {mitchell2021algorithmic}
\bibfield{author}{\bibinfo{person}{Shira Mitchell}, \bibinfo{person}{Eric Potash}, \bibinfo{person}{Solon Barocas}, \bibinfo{person}{Alexander D'Amour}, {and} \bibinfo{person}{Kristian Lum}.} \bibinfo{year}{2021}\natexlab{}.
\newblock \showarticletitle{Algorithmic fairness: Choices, assumptions, and definitions}.
\newblock \bibinfo{journal}{\emph{Annual Review of Statistics and Its Application}}  \bibinfo{volume}{8} (\bibinfo{year}{2021}), \bibinfo{pages}{141--163}.
\newblock


\bibitem[Ochigame(2020)]%
        {longhistory2020}
\bibfield{author}{\bibinfo{person}{Rodrigo Ochigame}.} \bibinfo{year}{2020}\natexlab{}.
\newblock \showarticletitle{The Long History of Algorithmic Fairness}.
\newblock \bibinfo{journal}{\emph{Phenomenal World}} (\bibinfo{year}{2020}).
\newblock


\bibitem[Penikas et~al\mbox{.}(2021)]%
        {penikas2021textual}
\bibfield{author}{\bibinfo{person}{Henry Penikas}, \bibinfo{person}{EA Fedorova}, \bibinfo{person}{AR Nevredinov}, {and} \bibinfo{person}{SM Druchok}.} \bibinfo{year}{2021}\natexlab{}.
\newblock \showarticletitle{Textual analysis of moral components in Islamic and Non-Islamic business in Russia}. In \bibinfo{booktitle}{\emph{2021 International Conference on Sustainable Islamic Business and Finance}}. IEEE, \bibinfo{pages}{140--143}.
\newblock


\bibitem[Rahwan(2018)]%
        {rahwan2018society}
\bibfield{author}{\bibinfo{person}{Iyad Rahwan}.} \bibinfo{year}{2018}\natexlab{}.
\newblock \showarticletitle{Society-in-the-loop: programming the algorithmic social contract}.
\newblock \bibinfo{journal}{\emph{Ethics and information technology}} \bibinfo{volume}{20}, \bibinfo{number}{1} (\bibinfo{year}{2018}), \bibinfo{pages}{5--14}.
\newblock


\bibitem[Ranalli and Malcom(2023)]%
        {ranalli2023s}
\bibfield{author}{\bibinfo{person}{Christopher Ranalli} {and} \bibinfo{person}{Finlay Malcom}.} \bibinfo{year}{2023}\natexlab{}.
\newblock \showarticletitle{What’s so bad about echo chambers?}
\newblock \bibinfo{journal}{\emph{Inquiry}} (\bibinfo{year}{2023}), \bibinfo{pages}{1--43}.
\newblock


\bibitem[Renner(2014)]%
        {renner2014globalization}
\bibfield{author}{\bibinfo{person}{Walter Renner}.} \bibinfo{year}{2014}\natexlab{}.
\newblock \showarticletitle{Globalization and Indian youth: Findings from moral foundations theory}. In \bibinfo{booktitle}{\emph{Current issues of science and research in the global world: Proceedings of the International Conference on Current Issues of Science and Research in the Global World, Vienna, Austria; 27}}. \bibinfo{pages}{7}.
\newblock


\bibitem[Romig et~al\mbox{.}(2018)]%
        {romig2018using}
\bibfield{author}{\bibinfo{person}{Charles~A Romig}, \bibinfo{person}{Virginia~T Holeman}, {and} \bibinfo{person}{Jill~Duba Sauerheber}.} \bibinfo{year}{2018}\natexlab{}.
\newblock \showarticletitle{Using moral foundations theory to enhance multicultural competency}.
\newblock \bibinfo{journal}{\emph{Counseling and Values}} \bibinfo{volume}{63}, \bibinfo{number}{2} (\bibinfo{year}{2018}), \bibinfo{pages}{180--193}.
\newblock


\bibitem[Sambasivan et~al\mbox{.}(2021)]%
        {sambasivan2021re}
\bibfield{author}{\bibinfo{person}{Nithya Sambasivan}, \bibinfo{person}{Erin Arnesen}, \bibinfo{person}{Ben Hutchinson}, \bibinfo{person}{Tulsee Doshi}, {and} \bibinfo{person}{Vinodkumar Prabhakaran}.} \bibinfo{year}{2021}\natexlab{}.
\newblock \showarticletitle{Re-imagining algorithmic fairness in india and beyond}. In \bibinfo{booktitle}{\emph{Proceedings of the 2021 ACM conference on fairness, accountability, and transparency}}. \bibinfo{pages}{315--328}.
\newblock


\bibitem[Selbst et~al\mbox{.}(2019)]%
        {selbst2019fairness}
\bibfield{author}{\bibinfo{person}{Andrew~D Selbst}, \bibinfo{person}{Danah Boyd}, \bibinfo{person}{Sorelle~A Friedler}, \bibinfo{person}{Suresh Venkatasubramanian}, {and} \bibinfo{person}{Janet Vertesi}.} \bibinfo{year}{2019}\natexlab{}.
\newblock \showarticletitle{Fairness and abstraction in sociotechnical systems}. In \bibinfo{booktitle}{\emph{Proceedings of the conference on fairness, accountability, and transparency}}. \bibinfo{pages}{59--68}.
\newblock


\bibitem[Shweder et~al\mbox{.}(2013)]%
        {shweder2013big}
\bibfield{author}{\bibinfo{person}{Richard~A Shweder}, \bibinfo{person}{Nancy~C Much}, \bibinfo{person}{Manamohan Mahapatra}, {and} \bibinfo{person}{Lawrence Park}.} \bibinfo{year}{2013}\natexlab{}.
\newblock \showarticletitle{The “big three” of morality (autonomy, community, divinity) and the “big three” explanations of suffering}.
\newblock In \bibinfo{booktitle}{\emph{Morality and health}}. \bibinfo{publisher}{Routledge}, \bibinfo{pages}{119--169}.
\newblock


\bibitem[Sivaraman et~al\mbox{.}(2023)]%
        {sivaraman2023ignore}
\bibfield{author}{\bibinfo{person}{Venkatesh Sivaraman}, \bibinfo{person}{Leigh~A Bukowski}, \bibinfo{person}{Joel Levin}, \bibinfo{person}{Jeremy~M Kahn}, {and} \bibinfo{person}{Adam Perer}.} \bibinfo{year}{2023}\natexlab{}.
\newblock \showarticletitle{Ignore, trust, or negotiate: understanding clinician acceptance of AI-based treatment recommendations in health care}. In \bibinfo{booktitle}{\emph{Proceedings of the 2023 CHI Conference on Human Factors in Computing Systems}}. \bibinfo{pages}{1--18}.
\newblock


\bibitem[Smith-Doerr et~al\mbox{.}(2017)]%
        {smith2017diversity}
\bibfield{author}{\bibinfo{person}{Laurel Smith-Doerr}, \bibinfo{person}{Sharla~N Alegria}, {and} \bibinfo{person}{Timothy Sacco}.} \bibinfo{year}{2017}\natexlab{}.
\newblock \showarticletitle{How diversity matters in the US science and engineering workforce: A critical review considering integration in teams, fields, and organizational contexts}.
\newblock \bibinfo{journal}{\emph{Engaging Science, Technology, and Society}}  \bibinfo{volume}{3} (\bibinfo{year}{2017}), \bibinfo{pages}{139--153}.
\newblock


\bibitem[Starke et~al\mbox{.}(2022)]%
        {starke2022fairness}
\bibfield{author}{\bibinfo{person}{Christopher Starke}, \bibinfo{person}{Janine Baleis}, \bibinfo{person}{Birte Keller}, {and} \bibinfo{person}{Frank Marcinkowski}.} \bibinfo{year}{2022}\natexlab{}.
\newblock \showarticletitle{Fairness perceptions of algorithmic decision-making: A systematic review of the empirical literature}.
\newblock \bibinfo{journal}{\emph{Big Data \& Society}} \bibinfo{volume}{9}, \bibinfo{number}{2} (\bibinfo{year}{2022}), \bibinfo{pages}{20539517221115189}.
\newblock


\bibitem[Telkamp and Anderson(2022)]%
        {telkamp2022implications}
\bibfield{author}{\bibinfo{person}{Jake~B Telkamp} {and} \bibinfo{person}{Marc~H Anderson}.} \bibinfo{year}{2022}\natexlab{}.
\newblock \showarticletitle{The implications of diverse human moral foundations for assessing the ethicality of Artificial Intelligence}.
\newblock \bibinfo{journal}{\emph{Journal of Business Ethics}} \bibinfo{volume}{178}, \bibinfo{number}{4} (\bibinfo{year}{2022}), \bibinfo{pages}{961--976}.
\newblock


\bibitem[Valdivia et~al\mbox{.}(2022)]%
        {valdivia2022there}
\bibfield{author}{\bibinfo{person}{Ana Valdivia}, \bibinfo{person}{J{\'u}lia~Corbera Serraj{\`o}rdia}, {and} \bibinfo{person}{Aneta Swianiewicz}.} \bibinfo{year}{2022}\natexlab{}.
\newblock \showarticletitle{There is an elephant in the room: Towards a critique on the use of fairness in biometrics}.
\newblock \bibinfo{journal}{\emph{AI and Ethics}} (\bibinfo{year}{2022}), \bibinfo{pages}{1--16}.
\newblock


\bibitem[Velasquez et~al\mbox{.}(1990)]%
        {velasquez1990justice}
\bibfield{author}{\bibinfo{person}{Manuel Velasquez}, \bibinfo{person}{Claire Andre}, \bibinfo{person}{T Shanks}, {and} \bibinfo{person}{Michael~J Meyer}.} \bibinfo{year}{1990}\natexlab{}.
\newblock \showarticletitle{Justice and fairness}.
\newblock \bibinfo{journal}{\emph{Issues in Ethics}} \bibinfo{volume}{3}, \bibinfo{number}{2} (\bibinfo{year}{1990}), \bibinfo{pages}{1--3}.
\newblock


\bibitem[Wenzel et~al\mbox{.}(2023)]%
        {wenzel2023can}
\bibfield{author}{\bibinfo{person}{Kimi Wenzel}, \bibinfo{person}{Nitya Devireddy}, \bibinfo{person}{Cam Davison}, {and} \bibinfo{person}{Geoff Kaufman}.} \bibinfo{year}{2023}\natexlab{}.
\newblock \showarticletitle{Can Voice Assistants Be Microaggressors? Cross-Race Psychological Responses to Failures of Automatic Speech Recognition}. In \bibinfo{booktitle}{\emph{Proceedings of the 2023 CHI Conference on Human Factors in Computing Systems}}. \bibinfo{pages}{1--14}.
\newblock


\end{thebibliography}

\end{document}